\documentclass[aps,pra,preprint,showpacs,showkeys]{revtex4}
\usepackage{amssymb,bm}
\usepackage{graphicx}
\usepackage{amsmath}
\usepackage{epstopdf}
\usepackage{float}
\allowdisplaybreaks

\begin{document}

\title{Coulomb effects in high-energy $e^+e^-$ electroproduction  by a heavy charged particles in an atomic field}

\author{P. A. Krachkov}\email{P.A.Krachkov@inp.nsk.su}
\author{A. I. Milstein}\email{A.I.Milstein@inp.nsk.su}
\affiliation{Budker Institute of Nuclear Physics, 630090 Novosibirsk, Russia}

\date{\today}

\begin{abstract}
The  cross section of high-energy $e^+e^-$ pair production by a  heavy charged particle in the atomic field is investigated in detail.
We take into  account  the interaction with the atomic field of $e^+e^-$ pair and a heavy particle as well. The calculation is performed exactly in the parameters of the atomic field. It is shown that, in contrast to the commonly accepted  point of view,  the cross section differential with respect to the final momentum of a heavy particle is  strongly affected by the interaction of a heavy particle with the atomic field. However, the cross section integrated over the final momentum of a heavy particle is  independent of this interaction.
\end{abstract}



\maketitle

\section{Introduction}
Production of   $e^+e^-$ pair  by the ultra-relativistic  heavy charged particle in the atomic field is very important because the cross section of this process is even larger than the cross section
of bremsstrahlung of a  heavy  particle in the field.  Thus, this process plays an important role in the energy losses of  heavy particles in detectors. The cross section of the process under consideration in the leading Born approximation was derived many years ago  \cite{Bhabha2,Racah37}. In this approximation, the cross section depends on the atomic charge number $Z$ and the charge number of a heavy particle $Z_p$ as $Z^2Z_p^2$. In  papers \cite{Bhabha2,Racah37} the interaction of a heavy particle with  the atomic field was not taken into account. Later, the Coulomb corrections with respect to the interaction of
$e^+e^-$ pair with the atomic field were obtained in Refs.~\cite{Nikishov82,IKSS1998} using the plane waves for the wave functions of a heavy particle and the Coulomb  wave functions for electron and positron. Thus, the results in Refs.~\cite{Nikishov82,IKSS1998} were exact in the parameter $Z\alpha$ but still proportional to    $Z_p^2$, where $\alpha$ is the fine-structure constant. The authors
of  Refs.~\cite{Nikishov82,IKSS1998}  obtained the cross sections differential with respect to the final momentum of a heavy particle as well as that integrated over this momentum. For the latter case, the cross sections were obtained within another approach, see Refs.~\cite{ISS1999,LM2000} and reviews \cite{BHT2007, H2008}. In that approach, the cross sections were obtained for a fixed impact parameter $\rho$ of a heavy particle with respect to the atomic center. Therefore, the interaction of a heavy particle with the atomic field was not taken into account. Then, the results were integrated over the impact parameter. Thus, the final results correspond to the cross sections integrated over the final momentum of a heavy particle. Note that the  energy $\omega$ of created $e^+e^-$ pair, which gives the main
contribution to the cross section,  is  much smaller than the energy of a heavy particle. As a result, the cross sections of the process are  independent of the spin and mass $m_p$ of a heavy particle but depend on the relativistic factor $\gamma=\varepsilon_p/m_p$, where  $\varepsilon_p$ is the energy of a heavy particle, $\hbar=c=1$.  Thus, the
formulas for the  cross sections are the same for muons and light nuclei (with the corresponding substitutions of the charge numbers).

In the present paper, we investigate  the differential cross section of  high-energy $e^+e^-$ electroproduction  by heavy charged particles in the  atomic field taking into account the interaction of a heavy particle with the atomic field. Our consideration is based on the quasiclassical approximation, see review in Ref.~\cite{KLM2016},  developed  in Ref.~\cite{KM2016} for the problem of high-energy  $e^+e^-$ electroproduction by ultra-relativistic  electron in the atomic field. Our results are exact in both parameters, $\eta=Z\alpha$ and $\eta_p=ZZ_p\alpha=Z_p\eta$. We show that the cross section differential over momentum of a heavy particle strongly depends on $\eta_p$, in contrast to the commonly accepted point of view. For light nuclei in the field of a heavy atom, this parameter can be large, $\eta_p\gtrsim1$.
However, the cross section integrated over the final momentum of a heavy particle is independent of the parameter $\eta_p$. It seems, the experimental investigation of a strong dependence of the differential cross section on  $\eta_p$ for moderate values of the relativistic factor $\gamma$ is not a very difficult task.

\section{General discussion}\label{general}

\begin{figure}[H]
\centering
\includegraphics[width=0.4\linewidth]{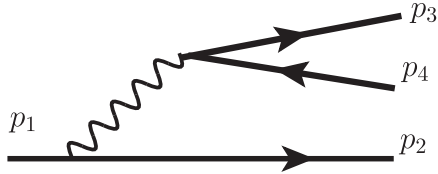}
\caption{Diagram for the amplitude of electroproduction by a heavy particle in the atomic field. Wavy line denotes the photon propagator, straight lines denote the wave functions in the atomic field.}
\label{fig:diagrams}
\end{figure}

The differential cross section of the process under consideration  reads~\cite{BLP1982}, see Fig.~\ref{fig:diagrams} where the corresponding Feynman diagrams in the Furry representation is shown:
\begin{equation}\label{eq:cs}
d\sigma=\frac{(Z_p\alpha)^2}{(2\pi)^8}\,d\varepsilon_3d\varepsilon_4\,d\bm p_{2\perp}\,d\bm p_{3\perp}d\bm p_{4\perp}\,|T|^{2}\,,
\end{equation}
where  $\bm p_1$ and $\bm p_2$ are  the  initial and final momenta of a heavy particle, $\bm p_3$ and  $\bm p_4$ are the momenta of electron and positron, $\varepsilon_{1}=\varepsilon_{2}+\omega$ is the  energy of the  incoming heavy particle, $\omega=\varepsilon_{3}+\varepsilon_{4}$,  $\varepsilon_{1,2}=\sqrt{{p}_{1,2}^2+m^2_p}$, $\varepsilon_{3,4}=\sqrt{{p}_{3,4}^2+m^2_e}$, $m_e$ is the electron mass and $m_p$ is the mass of a heavy particle.  In Eq.~\eqref{eq:cs} the notation  $\bm X_\perp=\bm X-(\bm X\cdot\bm \nu)\bm\nu$ for any vector $\bm X$ is used, $\bm\nu=\bm p_1/p_1$. Below  we assume that $\varepsilon_{1,2}\gg m_p$, $\varepsilon_{3,4}\gg m_e$, and
$m_p\gg m_e$. The main contribution to the cross section is given by the energy region $\varepsilon_{3,4}\lesssim \gamma m_e$ , where $\gamma=\varepsilon_{1}/m_p\gg 1$, so that
$\omega/\varepsilon_{1}\lesssim m_e/m_p\ll 1$.

The straightforward application of the method developed in our recent paper \cite{KM2016} results in the expression for $T$ exact in the parameters $\eta$ and $\eta_p$ for arbitrary atomic potential $V(r)$. We perform  calculations of the matrix element $T$  for definite helicities $\mu_3$ and $\mu_4$ of electron and positron, respectively, $\mu_i=\pm$  denotes  a  sign of the helicity.

 We begin  our consideration with the case of a pure Coulomb field and then discuss the effects of screening.

\subsection{Coulomb field}\label{CF}

It is convenient to write the amplitude $T$ as a sum $T=T_\perp+T_\parallel$, where the corresponding contributions have the form (cf. Eq.~(18) in Ref.~\cite{KM2016}):
\begin{align}\label{T1C}
&T_\perp=\frac{8i\eta}{\omega}|\Gamma(1-i\eta)|^2 \int\frac{d\bm\Delta_\perp\, A_{as}(\bm\Delta_\perp+\bm p_{2\perp})}{(Q^2+\Delta_{0\parallel}^2) M^2\,(\omega^2/\gamma^2+\Delta_{\perp}^2)}\left(\frac{\xi_2}{\xi_1}\right)^{i\eta}
{\cal M}\,, \nonumber\\
&{\cal M}=-\frac{\delta_{\mu_3\bar\mu_4}}{\omega} \big[ \varepsilon_3(\bm s_{\mu_3}^*\cdot\bm\Delta_{\perp})(\bm s_{\mu_3}\cdot\bm I_1)
-\varepsilon_4(\bm s_{\mu_4}^*\cdot\bm\Delta_{\perp})(\bm s_{\mu_4}\cdot\bm I_1)  \big]
+\mu_3\delta_{\mu_3\mu_4}\frac{m_e}{\sqrt{2}}(\bm s_{\mu_3}^*\cdot\bm\Delta_{\perp})I_0\,,\nonumber\\
&T_\parallel=-\frac{8i\eta\varepsilon_3\varepsilon_4}{\omega^3}|\Gamma(1-i\eta)|^2 \int \frac{d\bm\Delta_\perp\, A_{as}(\bm\Delta_\perp+\bm p_{2\perp})}{(Q^2+\Delta_{0\parallel}^2) M^2}\left(\frac{\xi_2}{\xi_1}\right)^{i\eta}\,I_0
\delta_{\mu_3\bar\mu_4}\,,
\end{align}
where $\omega=\varepsilon_3+\varepsilon_4$,  $\bm s_{\lambda}=(\bm e_x+i\lambda\bm e_y)/\sqrt{2}$,
 $\bm e_x$ and  $\bm e_y$ are two orthogonal unit vectors perpendicular to $\bm\nu=\bm p_1/p_1$, $\Gamma(x)$ is the Euler $\Gamma$ function, and the function  $A_{as}(\bm\Delta_\perp)$ reads
 \begin{align}\label{ACJCas}
&A_{as}(\bm\Delta_\perp)=-\frac{4\pi\eta_p (L\Delta_\perp)^{2i\eta_p}\Gamma(1-i\eta_p)}{\Delta_\perp^2\Gamma(1+i\eta_p)}\,.
\end{align}
A specific value of $L$ is irrelevant because the factor $L^{2i\eta_p}$ disappears in $|T|^2$.
In Eq.~\eqref{T1C} the following notations are used:
\begin{align}\label{T1Cnot}
&\bm\Delta_{0\perp}=\bm p_{2\perp}+\bm p_{3\perp}+\bm p_{4\perp}\,,\quad
\Delta_{0\parallel}=-\frac{1}{2}\left[\frac{\omega}{\gamma^2}+\frac{\omega(m_e^2+\zeta^2)}{\varepsilon_3\varepsilon_4}+\frac{\delta^2}{\omega}+
\frac{p_{2\perp}^2}{\varepsilon_1}\right]\,,\nonumber\\
&M^2=m^2_e+\frac{\varepsilon_3\varepsilon_4}{\gamma^2}
+\frac{\varepsilon_3\varepsilon_4}{\omega^2}\Delta_{\perp}^2\,,\quad
 \bm\zeta=\frac{\varepsilon_4}{\omega}\bm p_{3\perp}-\frac{\varepsilon_3}{\omega}\bm p_{4\perp}\,,\quad \bm\delta=\bm p_{3\perp}+\bm p_{4\perp}\,,\nonumber\\
&\bm Q=\bm\Delta_\perp-\bm\delta\,,\quad
\bm q_1=\frac{\varepsilon_3}{\omega}\bm Q- \bm\zeta\,,\quad \bm q_2=
 \frac{\varepsilon_4}{\omega}\bm Q+ \bm\zeta\,,\nonumber\\
&I_0=(\xi_1-\xi_2)F(x)+(\xi_1+\xi_2-1)(1-x)\frac{F'(x)}{i\eta}\,,\nonumber\\
&\bm I_1=(\xi_1\bm q_1+\xi_2\bm q_2)F(x)+(\xi_1\bm q_1-\xi_2\bm q_2)(1-x)\frac{F'(x)}{i\eta}\,,\nonumber\\
&\xi_1=\frac{M^2}{M^2+q_1^2}\,,\quad \xi_2=\frac{M^2}{M^2+q_2^2}\,,\quad x=1-\frac{Q^2\xi_1\xi_2}{M^2}\,,\nonumber\\
&F(x)=F(i\eta,-i\eta, 1,x)\,,\quad F'(x)=\frac{\partial}{\partial x}F(x)\,,
\end{align}
where  $F(a,b,c,x)$ is the hypergeometric function.

In terms of the variables $\bm\zeta$, $\bm\delta$, and  $\bm p_{2\perp}$, see  Eq.~\eqref{T1Cnot},
the main contribution to the total cross section is given by the region $\zeta,\, \delta,\,p_{2\perp}\lesssim m_e$ and $\omega\lesssim m_e\gamma$.
In this region  $\omega/\varepsilon_1\lesssim m_e/m_p$ and  $p_{2\perp}/\varepsilon_1\ll p_{3\perp}/\varepsilon_3,\,p_{4\perp}/\varepsilon_4$, i.e., the angle between the momenta $\bm p_2$ and $\bm p_1$
is much smaller than the angles between $\bm p_3$, $\bm p_4$ and $\bm p_1$.

Let us first  consider  the cross section integrated over $\bm p_{2\perp}$, which is one of the most interesting quantity from the experimental point of view. We show that this cross section is independent of the parameter $\eta_p$,  so that  it is not affected by the  interaction
of a heavy particle with the Coulomb center.  First of all we note that the last term   $ p_{2\perp}^2/\varepsilon_1$  in  $\Delta_{0\parallel}$ (see Eq.~\eqref{T1Cnot}) may be omitted because its contribution is small as compared with that of other terms (the relative contribution of this term to $\Delta_{0\parallel}$ is $m_e/m_p$). After that the variable $\bm p_{2\perp}$ presents in the amplitude $T$  solely in the function   $A_{as}(\bm\Delta_\perp+\bm p_{2\perp})$, see Eq.~\eqref{T1C}.    Let us consider the integral,
\begin{equation}\label{eq:examp}
R=\int\,d \bm p_{2\perp}\,\left|\int d\bm\Delta_{\perp} A_{as}(\bm\Delta_\perp+\bm p_{2\perp})G(\bm\Delta_\perp)\right |^{2}\,,
\end{equation}
where $G(\bm\Delta_\perp)$ is some function and  $A_{as}(\bm\Delta_\perp)$ is given by Eq.~\eqref{ACJCas}. The integral \eqref{eq:examp} is well-defined. However, to have a possibility to  change  the order of integration, we introduce the regularized function $A_{as}^{(r)}(\bm\Delta_\perp)$,
\begin{align}\label{ACJCasr}
&A_{as}^{(r)}(\bm\Delta)=-\frac{4\pi(\eta_p-i\epsilon) (L\Delta_\perp)^{2i\eta_p+2\epsilon}\Gamma(1-i\eta_p-\epsilon)}{\Delta_\perp^2\Gamma(1+i\eta_p+\epsilon)}\,,
\end{align}
where $\epsilon$ is a small positive parameter of regularization.
Then we have
\begin{equation}\label{eq:exampreg}
R=\lim_{\epsilon\rightarrow 0}\iint d\bm x  d\bm y\, G(\bm x)G^*(\bm y) \int\,d \bm p_{2\perp}\, A_{as}^{(r)}(\bm x+\bm p_{2\perp})
A_{as}^{(r)*}(\bm y+\bm p_{2\perp})\,.
\end{equation}
The integral over $\bm p_{2\perp}$ can be easily take by means of the Feynman parametrization. We have
\begin{equation}\label{eq:exampreg1}
R=\lim_{\epsilon\rightarrow 0}\iint d\bm x  d\bm y\, G(\bm x)G^*(\bm y) \frac{32\pi^3\epsilon}{|\bm x-\bm y|^{2-4\epsilon}}=(2\pi)^4\int d\bm x |G(\bm x)|^2\,.
\end{equation}
Thus, the final result is independent of the parameter $\eta_p$, so that it can be evaluated in the limit $\eta_p\rightarrow 0$ using the relation
\begin{equation}
\lim_{\eta_p\rightarrow 0}\,\eta_p\int d\bm\Delta_\perp\,|\bm\Delta_\perp+\bm p_{2\perp}|^{2i\eta_p-2}G(\bm\Delta_\perp)=-i\pi G(-\bm p_{2\perp})\,.
\end{equation}
The cross section $d\sigma_0$ in this limit is given by Eq.~\eqref{eq:cs} with the replacement $T\rightarrow \cal T={\cal T}_\perp+{\cal T}_\parallel$ with
\begin{align}\label{calT}
&{\cal T}_\perp=-\frac{{32\pi^2\eta|\Gamma(1-i\eta)|^2 \cal M}_0}{\omega\Delta_{0}^2 M^2\,(\omega^2/\gamma^2+ p_{2\perp}^2)}\left(\frac{\xi_2}{\xi_1}\right)^{i\eta}
\,, \nonumber\\
&{\cal M}_0=\frac{\delta_{\mu_3\bar\mu_4}}{\omega} \big[ \varepsilon_3(\bm s_{\mu_3}^*\cdot\bm p_{2\perp})(\bm s_{\mu_3}\cdot\bm I_1)
-\varepsilon_4(\bm s_{\mu_4}^*\cdot\bm p_{2\perp})(\bm s_{\mu_4}\cdot\bm I_1)  \big]\,,\nonumber\\
&-\mu_3\delta_{\mu_3\mu_4}\frac{m_e}{\sqrt{2}}(\bm s_{\mu_3}^*\cdot\bm p_{2\perp})I_0\,,\nonumber\\
&{\cal T}_\parallel= \frac{32\pi^2\eta\varepsilon_3\varepsilon_4|\Gamma(1-i\eta)|^2 }{\omega^3\Delta_{0}^2 M^2}\left(\frac{\xi_2}{\xi_1}\right)^{i\eta}\,I_0
\delta_{\mu_3\bar\mu_4}\,,
\end{align}
where all notations are given in Eq.~\eqref{T1Cnot} with the replacement $\bm\Delta_\perp\rightarrow-\bm p_{2\perp}$. The result \eqref{calT} agrees with that obtained in Ref.~\cite{Nikishov82}.

\begin{figure}[h]
\centering
\includegraphics[width=0.7\linewidth]{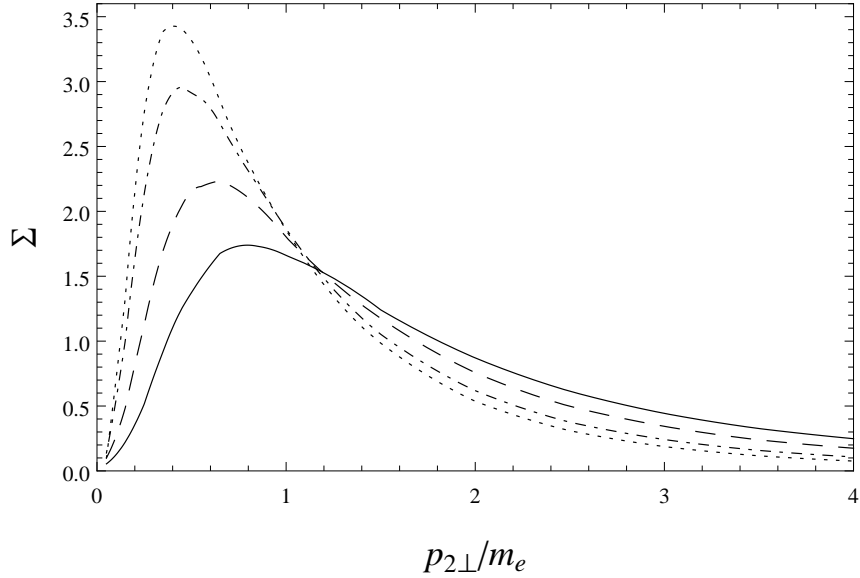}
\caption{The dependence of $\Sigma$, Eq.~\eqref{Sig},  on $p_{2\perp}/m_e$ for $\omega=m_e\gamma/4$, $\varepsilon_3=\varepsilon_4=\omega/2$,  $\gamma=100$, $Z=79$ (gold), and a few values of $Z_p$; solid curve for $Z_p=3$, dashed curve for  $Z_p=2$,  dash-dotted curve for $Z_p=1$, and  dotted curve for $Z_p\rightarrow 0$ (without account for the interaction of a heavy particle with the Coulomb field).}
\label{difpee}
\end{figure}

Though the cross section integrated over $\bm p_{2\perp}$ is independent of $\eta_p$,  the cross section differential over $\bm p_{2\perp}$ strongly depends on this parameter. This statement is illustrated in Fig.~\ref{difpee}, where the quantity $\Sigma$,
 \begin{equation}\label{Sig}
\Sigma=\frac{d\sigma}{Sdp_{2\perp}d\varepsilon_3d\varepsilon_4}\,, \quad S=\frac{(Z_p\alpha)^2}{\omega^2m_e^3}\,,
\end{equation}
which is the differential cross section in units $S$ integrated over $\bm p_{3\perp}$ and $\bm p_{4\perp}$, is shown as the function of  $p_{2\perp}$ for
$\omega=m_e\gamma/4$, $\varepsilon_3=\varepsilon_4=\omega/2$,  $\gamma=100$, $Z=79$ (gold), and a few values of $Z_p$.

It is seen that  impact of interaction of a heavy particle with the Coulomb field on the cross section differential over $\bm p_{2\perp}$ is  significant.
At small value of  $p_{2\perp}/m_e$,  the cross section exact in $\eta_p$  is essentially smaller than that obtained in the limit  $\eta_p\rightarrow 0$. At large value of $p_{2\perp}/m_e$, the relation between these cross sections is opposite. As should be, $\int_0^\infty \Sigma\, dp_{2\perp}$  is independent of $\eta_p$.

\begin{figure}[h]
\centering
\includegraphics[width=0.45\linewidth]{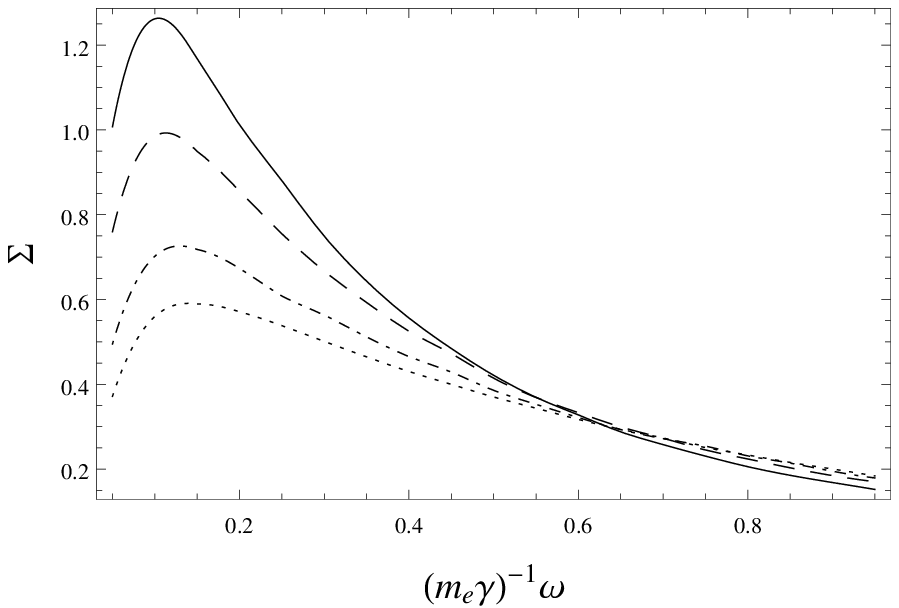}
\includegraphics[width=0.45\linewidth]{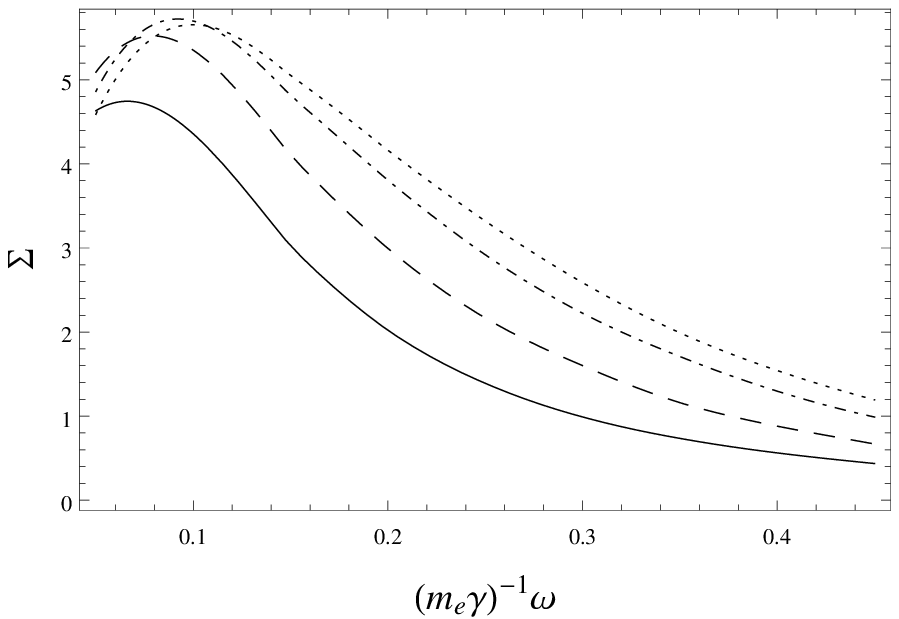}
\caption{The dependence of $\Sigma$, Eq.~\eqref{Sig},  on $\omega/(m_e\gamma)$ for $p_{2\perp}/m_e=2$ (left picture) and $p_{2\perp}/m_e=0.5$ (right picture), $\varepsilon_3=\varepsilon_4$,  $\gamma=100$, $Z=79$ (gold), and a few values of $Z_p$; solid curve for $Z_p=3$, dashed curve for  $Z_p=2$,  dash-dotted curve for $Z_p=1$, and  dotted curve for $Z_p\rightarrow 0$ (without account for the interaction of a heavy particle with the Coulomb field).}
\label{difom}
\end{figure}

In Fig.~\ref{difom} the quantity $\Sigma$  is shown as the function of  $\omega/(m_e\gamma)$ for
 $p_{2\perp}/m_e=2$ (left picture) and $p_{2\perp}/m_e=0.5$ (right picture), $\varepsilon_3=\varepsilon_4$,  $\gamma=100$, $Z=79$ (gold), and a few values of $Z_p$. It is seen that the dependence of the function $\Sigma$ on $Z_p$ is very strong for all  $\omega/(m_e\gamma)$.

\begin{figure}[h]
\centering
\includegraphics[width=0.4\linewidth]{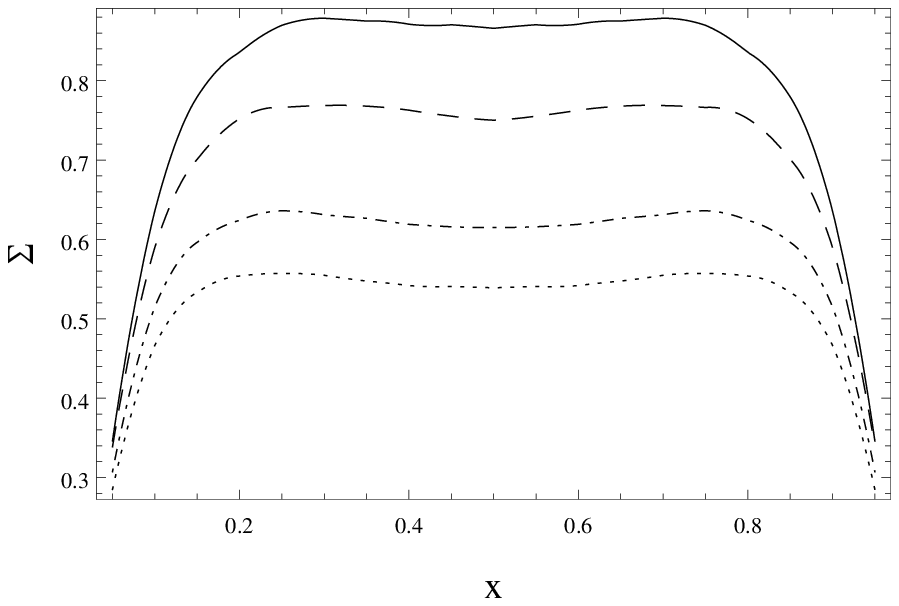}
\includegraphics[width=0.4\linewidth]{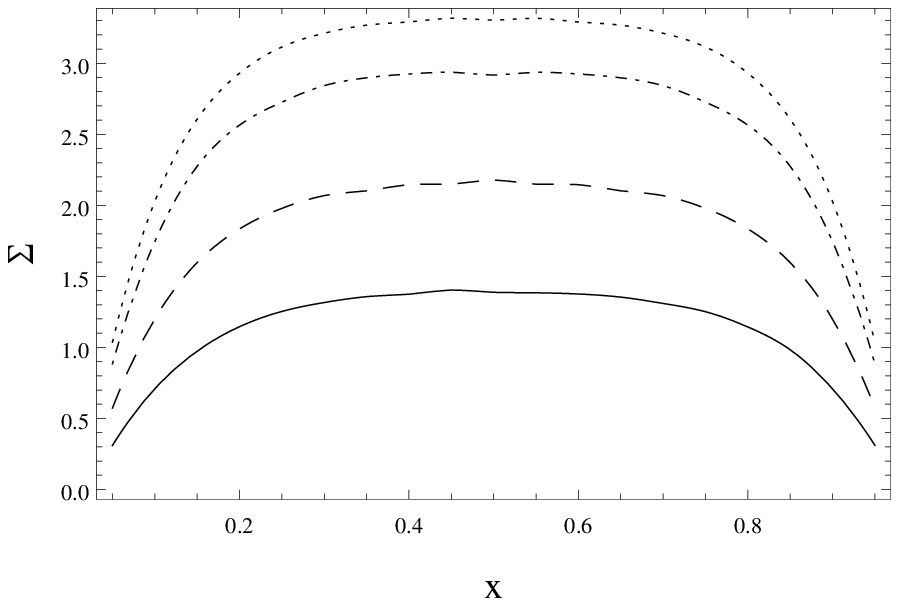}
\caption{The dependence of $\Sigma$, Eq.~\eqref{Sig},  on $x=\varepsilon_3/\omega$ for  $\omega=m_e\gamma/4$,  $\gamma=100$, $Z=79$ (gold), $p_{2\perp}/m_e=2$ (left picture), $p_{2\perp}/m_e=0.5$ (right picture), and a few values of $Z_p$; solid curve for $Z_p=3$, dashed curve for  $Z_p=2$,  dash-dotted curve for $Z_p=1$, and  dotted curve for $Z_p\rightarrow 0$ (without account for the interaction of a heavy particle with the Coulomb field).}
\label{difx}
\end{figure}

In Fig.~\ref{difx} we show the dependence of $\Sigma$  on $x=\varepsilon_3/\omega$  for  $\omega=m_e\gamma/4$, $\gamma=100$, $Z=79$ (gold), $p_{2\perp}/m_e=2$ (left picture), $p_{2\perp}/m_e=0.5$ (right picture), and a few values of $Z_p$. Again, it is seen that the account for the interaction of a heavy particle with the Coulomb field is very important for the differential cross section.

\subsection{Effect of screening}\label{Scr}
  An account for the effect of screening  can be performed in the same way as it has been done in our paper \cite{KM2016}. In Eq.~(\ref{T1C}) one should  replace $A_{as}(\bm\Delta_\perp)\rightarrow A(\bm\Delta_\perp)$,
\begin{align}\label{eq:Aperp}
&A(\bm\Delta_\perp)=i\int d\bm\rho\,\exp[-i\bm\Delta_\perp\cdot\bm\rho-iZ_p\chi(\rho)]\,,\nonumber\\
&\chi(\rho)=\int_{-\infty}^\infty dz\,V(\sqrt{z^2+\rho^2})\,,
\end{align}
where $V(r)$ is the atomic potential,  and multiply the integrand in Eq.~(\ref{T1C})  by the function $F_a((\bm\Delta_\perp-\bm\delta)^2+\Delta_{0\parallel}^2)$, with $F_a(q^2)$  being  the atomic form factor.

 We show that the cross section integrated over $\bm p_{2\perp}$ is independent of $\eta_p$ for any localized potential $V(r)$. Let us consider the function $R_1$,
\begin{equation}\label{eq:examp1}
R_1=\int\,d \bm p_{2\perp}\,\left|\int d\bm\Delta_{\perp} A(\bm\Delta_\perp+\bm p_{2\perp})G(\bm\Delta_\perp)\right |^{2}\,.
\end{equation}
 Substituting Eq.~\eqref{eq:Aperp} to  Eq.~\eqref{eq:examp1} we obtain
 \begin{align}\label{eq:examp2}
&R_1=\iint d\bm x d\bm y G(\bm x)G^*(\bm y)\iint d\bm\rho_1d\bm\rho_2\exp\{iZ_p[\chi(\rho_2)-\chi(\rho_1)]+i\bm y\cdot\bm\rho_2-i\bm x\cdot\bm\rho_1\}\nonumber\\
&\times\int\,d\bm p_{2\perp}\exp[i\bm p_{2\perp}\cdot(\bm\rho_2-\bm\rho_1)]\,.
\end{align}
Taking the integrals first over $\bm p_{2\perp}$ and then over $\bm\rho_1$,  $\bm\rho_2$, and $\bm y$, we find the following result
\begin{equation}\label{eq:relR1}
R_1=(2\pi)^4\int d\bm p_{2\perp} |G(\bm p_{2\perp})|^2\,,
\end{equation}
which is independent of $\eta_p$. Thus, the cross section integrated over $\bm p_{2\perp}$ can be evaluated by means of Eq.~\eqref{eq:cs}, where $T=F_a(\Delta_0^2)({\cal T}_\perp+{\cal T}_\parallel)$,
${\cal T}_\perp$ and  ${\cal T}_\parallel$ are given in Eq.~\eqref{calT}.

Let us  discuss  the impact of screening on the differential  cross section.
Without account for  the interaction of a heavy particle with the atomic field, this question was investigated in Ref.~\cite{MUT}. In this case the effect of screening is important for
$\gamma\gg m_er_{scr}\sim Z^{-1/3}/\alpha$, where $r_{scr}$ is the screening radius. Screening modifies the result of account for  the interaction of a heavy particle with the atomic field in the cross section differential over $p_{2\perp}$ for $\gamma\gg (\omega/m_e)m_er_{scr}\gg m_er_{scr}$. Therefore, up to a very large $\gamma$  one  can use $A_{as}(\bm\Delta)$, Eq.~\eqref{ACJCas},   instead of
$A(\bm\Delta)$, Eq.~\eqref{eq:Aperp}, but take into account the atomic form factor $F_a$.

\section{Conclusion}
Using the quasiclassical approximation, we have derived the  differential cross section of high-energy $e^+e^-$ electroproduction by heavy charged particles in the atomic field. The result is exact in the parameters $\eta$ and $\eta_p$. It is shown  that the cross section differential in $\bm p_{2\perp}$ strongly depends on the parameter $\eta_p$ while the cross section integrated over $\bm p_{2\perp}$ is independent of this parameter. Though, at $\gamma \gtrsim m_er_{scr}\gg 1$, screening is important for interaction of $e^+e^-$ pair with the atomic field, it is not necessary to take  screening into account for interaction of a heavy particle with the atomic field up to a very large $\gamma$. 

Situation with the dependence of the cross sections  of electroproduction on $\eta_p$  reminds  the situation with the Coulomb corrections to the cross sections of bremsstrahlung by high-energy muons in the atomic field, see Ref.~\cite{KM2015}. The Coulomb corrections to the cross section of bremsstrahlung differential in both final muon  and photon momenta modify essentially the result as compared  with the Born result. However, the Coulomb corrections to the cross section integrated over the final muon momentum (or the photon momentum) vanish.

To observe experimentally a strong  dependence of the cross section under discussion on the value of the parameter $\eta_p$, it is necessary to detect the final heavy particle. The angle between vectors 
$\bm p_2$ and $\bm p_1$ is essentially  smaller than the angles between $\bm p_3$, $\bm p_4$, and $\bm p_1$.  However, this experiment for moderate values of the relativistic factor $\gamma$ seems to be   not a very hard problem.

\section*{Acknowledgement}
This work has been supported by Russian Science Foundation (Project No. 14-50-00080). It has been also supported in part by
RFBR (Grant No. 16-02-00103).

 \end{document}